\documentclass[referee]{aasstitre} 
\usepackage{epsfig}
\begin{document}
  \thesaurus{09.18.1} 
   \title{Is there ERE in  diffuse galactic light at high latitude?}

   \author{Fr\'ed\'eric Zagury \inst{}  }


   \institute{Department of Astrophysics, Nagoya University,
              Nagoya, 464-01 Japan 
              \\
              email: zagury@a.phys.nagoya-u.ac.jp
             }

   \date{January, 2000}
   

  \titlerunning{ERE in  diffuse galactic light }
  
  \maketitle

 \begin{abstract}
Up until recently most optical observations of nebulae were explained by 
scattering of starlight. 
Since the 1970s, authors have believed optical observations of 
interstellar matter couldn't be accounted for 
by extinction effects only. To join observations and models 
they required an additional broad band emission in the $R$ and $I$ band range. 
This emission, called ERE (extended red emission), is attributed to 
luminescence of some molecule of the interstellar medium which hasn't been identified yet.

In 1998 Gordon, Witt and Friedmann wrote of detection of
ERE in diffuse galactic light at high latitudes. 
The comparaison of diffuse galactic light and starlight over a few 
regions of the sky cannot, according to the authors, be fully explained by 
extinction of starlight.

Different aspects of Gordon et al.'s data reduction work, and of the 
model which supports their assertion, will be considered here.
IRAS images show that the scattering medium in which diffuse galactic 
light originates is misrepresented when using their model.
The quantity of matter on the stars' line of sight, 
used to calculate the $R$ magnitude of the 
stars, is in many cases underestimated, leading to a color of 
starlight bluer than the exact one.
Finally, the search for ERE could have been simplified by the comparison of 
starlight and Pioneer colors, which are equal within the error bars.

Analysis of the $100\,\mu$m emission of the regions Gordon et al. have studied will prove 
that the color of the diffuse galactic 
light is explained by scattering of the background starlight by 
dust embedded in nearby cirrus. 

ERE may not be present in diffuse galactic light, and, more 
generally, in interstellar space.

   \end{abstract} 
 \section{Introduction}
To study the optical properties of interstellar grains two 
types of approaches are usually followed. 
The first is to compare the strength of a nebula surface brightness
to the value of the source radiation field at the cloud position. 
More frequently the color of the cloud is compared to that of the  
illuminating source.

Both methods have led some authors (Lynds~\cite{lynds62}, 
Guhathakurta~\&~Cutri~\cite{gu2}, Gordon, Witt and Friedmann \cite{gordon} 
and references 
therein) to conclude that not all interstellar cloud emission can be accounted 
for solely by the scattering of starlight. 
According to these authors, comparison of nebula and source emissions 
cannot explain the strength of the emission in the $R$ and $I$ bands, 
and the red color  emission of some, if not most, interstellar clouds. 
Hence, the clouds must be emitting themselves, which implies the 
presence of a particular class 
of dust grain, assumed to absorb UV ambient radiation and to
re-emit in the red. The phenomenom is called ERE (Extended Red 
Emission). Unfortunately, all the attempts to identify ERE carriers have 
failed. 

The directions where ERE is found, which for many years consisted of only one 
direction (the Red 
Rectangle), now 
cover nearly the entire sky. The phenomenon is thought to be so important that 
up to $50{\%}$ of interstellar red emission is attributed to ERE (Gordon, 
Witt and Friedmann \cite{gordon}, hereafter GWF). ERE appears in a wide range of 
different 
environments and astrophysical objects, including cirrus, 
nebulae, planetary nebulae, HII 
regions, novae, and other galaxies. Complete references can be found in 
the introduction of GWF.

Diffuse galactic light (DGL) was identified at the beginning of the century (Struve~\& Elvey 
\cite{struve36}) as diffuse 
interstellar radiation which remains after the subtraction of direct 
starlight and the light of solar and terrestial origins.
It was attributed to starlight scattered by interstellar grains.
Toller (\cite{toller81}) has found that the visible 
surface brightness of the DGL in high 
latitude directions and the HI column density are roughly proportional.

The GWF paper is devoted to detecting ERE in  DGL and is divided into two parts.

The first part of the GWF paper describes the method the authors have 
followed to estimate the $B$ and $R$ surface brightnesses of direct starlight in 
different directions of the galaxy.
In two high galactic latitude regions, region (a) and region (b), 
the result is substracted from the $B$ and $R$ Pioneer satellite measurements 
to yield an estimate of the DGL.

The second part of the GWF paper (\S~4 on), in which the authors 
attempt to demonstrate the existence of ERE in the diffuse 
interstellar medium, is constructed on the following idea:
if we have a reliable model of the starlight scattered from the 
diffuse interstellar medium, and if the observed DGL is 
in excess of this estimate, the difference must arise from a 
non-scattering emission process.

To estimate the scattered part of DGL, GWF use the Witt-Peterson 
model (`WP' model hereafter) of the galaxy. 
From HI data, the model sets up a vast three dimensional representation of the galaxy
which GWF use to calculate 
the amount of scattered-DGL received from any direction.

The large grains responsible for the scattered optical DGL also 
have a thermal $100\,\mu$m emission.
Following Toller's search for an HI counterpart to the DGL, IRAS 
images are a powerful tool to visualize the medium in which the 
DGL originates.
For the regions of the sky in which GWF have separated DGL from 
starlight, IRAS images  can be used to check the pertinence and accuracy of 
the WP model.
In both regions, the medium has a complicated structure.
It is impossible for the WP model, or any other model, to approach the three dimensional 
structure of these mediums with the necessary precision solely from
an estimate of the HI column density.
It is surprising that the large differences between the 
results of the model and the observations are attributed either to 
variations of the properties of interstellar grains with longitude when the $B$ band 
is concerned, or to ERE for the $R$ band.

Is it even possible, or furthermore necessary, to model 
the part of the DGL which comes from the scattering of starlight?
Since the interstellar grains are forward scattering light 
(Henyey~\& Greenstein \cite{henyey}), 
DGL in one direction must arise from the scattering of the light of 
stars in close proximity to this direction.
Evaluation of the scattered part of the DGL can be restricted to a 
local estimate of the radiative transfer of the background direct 
starlight through the interstellar medium.
The radiative transfer does not need to be calculated for the entire galaxy. 
Existence of ERE could have been probed 
through a direct and local comparison of DGL and direct starlight.
Such a comparison (section~\ref{rad}) is more reliable since it will 
use GWF data only.

IRAS images help to understand the 
variation of the DGL and the starlight surface brightnesses
in fields (a) and (b), in relation to the interstellar medium structure.
There are clear correlations between IRAS $100\,\mu$m images and the 
variations of DGL and direct starlight in the two regions where it was 
isolated by GWF.
These correlations will be discussed in section~\ref{exp}.

The data presented in the GWF paper are reviewed in section~\ref{data}.
The difficulties which arise when considering the GWF method and the WP model
are detailed in section~\ref{gwfarg} and section~\ref{crit}.
An alternative explanation of the GWF data, supported by the IRAS 
images of fields (a) and (b), is proposed from section~\ref{rad} to 
section~\ref{exp}.

To conclude, all of GWF data will be 
explained by scattering of background starlight by the nearby 
interstellar cirrus.
The relative colors of Pioneer 
observations, the direct starlight and the DGL can be explained with 
the absence of ERE.

\clearpage
\begin{figure*}
\resizebox{\columnwidth }{!}{\includegraphics{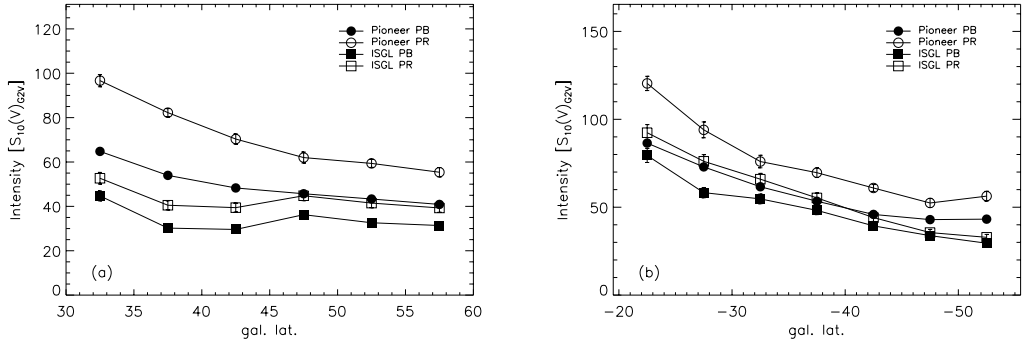}} 
\caption{GWF figure 12. Caption: `Red and blue 
intensities for the Pioneer measurements and integrated star/galaxy 
light (ISGL) are plotted as functions of Galactic latitude for cuts 
in Galactic longitude between $0^{\circ}$ and $5^{\circ}$ (a) and 
$95^{\circ} $ and  $100^{\circ}  $ (b). Each point corresponds to a $ 5^{\circ} \times
5^{\circ}$ region. The Pioneer error bars were computed using the 
algorithm described in \S 2.1. The ISGL error bars were assumed to be 
a conservative $ 5\% $ (see \S 3.2).'
}
\label{gord12}
\end{figure*}
\nopagebreak
\begin{figure*}
\resizebox{\columnwidth }{!}{\includegraphics{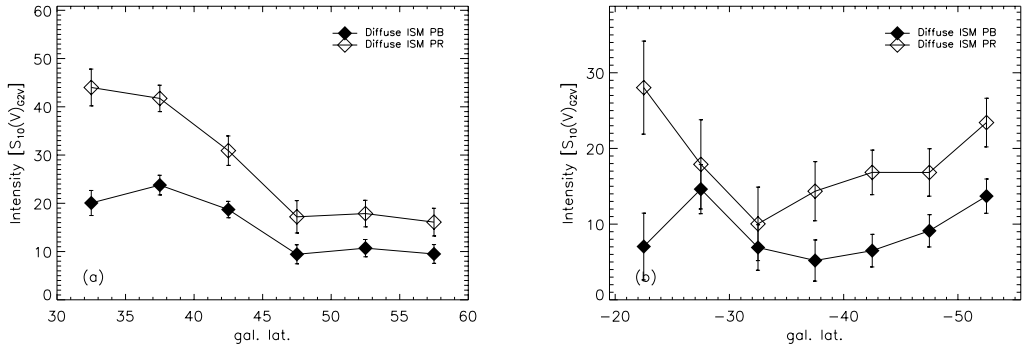}} 
\caption{GWF figure 13. Caption: `Red and blue 
intensities for the light from the diffuse ISM are plotted for cuts 
in Galactic longitude between  $0^{\circ}$ and  $5^{\circ}$ 
(a) and  $95^{\circ}$ and  $100^{\circ}$ (b). The diffuse ISM 
intensities were computed by subtracting the ISGL from the Pioneer
measurements.'} 
\label{gord13}
\end{figure*}
\nopagebreak
\begin{figure*}
\resizebox{\columnwidth }{!}{\includegraphics{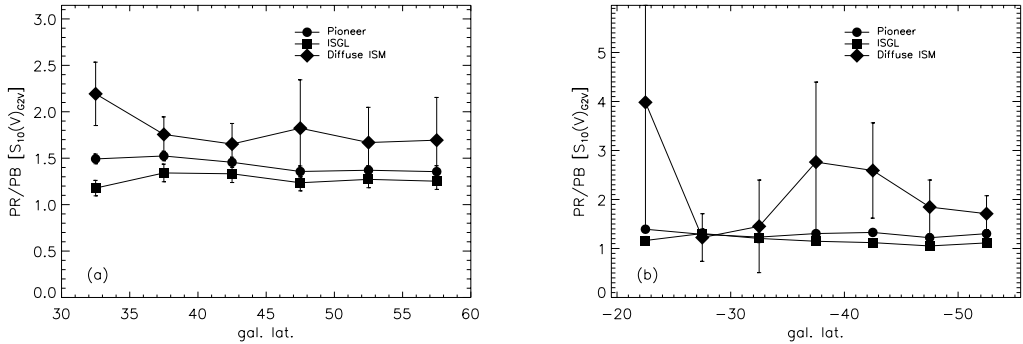}} 
\caption{GWF figure 14. Caption: `Red/blue ratio 
for the Pioneer measurements, the ISGL, and the diffuse ISM. The 
first plot (a) displays the cut in Galactic longitude between  
$0^{\circ}$ and $ 5^{\circ} $ and the second plot (b), the cut 
between  $95^{\circ}$ and $ 100^{\circ}$.'} 
\label{gord14}
\end{figure*}
\clearpage
\begin{figure}
\resizebox{\columnwidth }{!}{\includegraphics{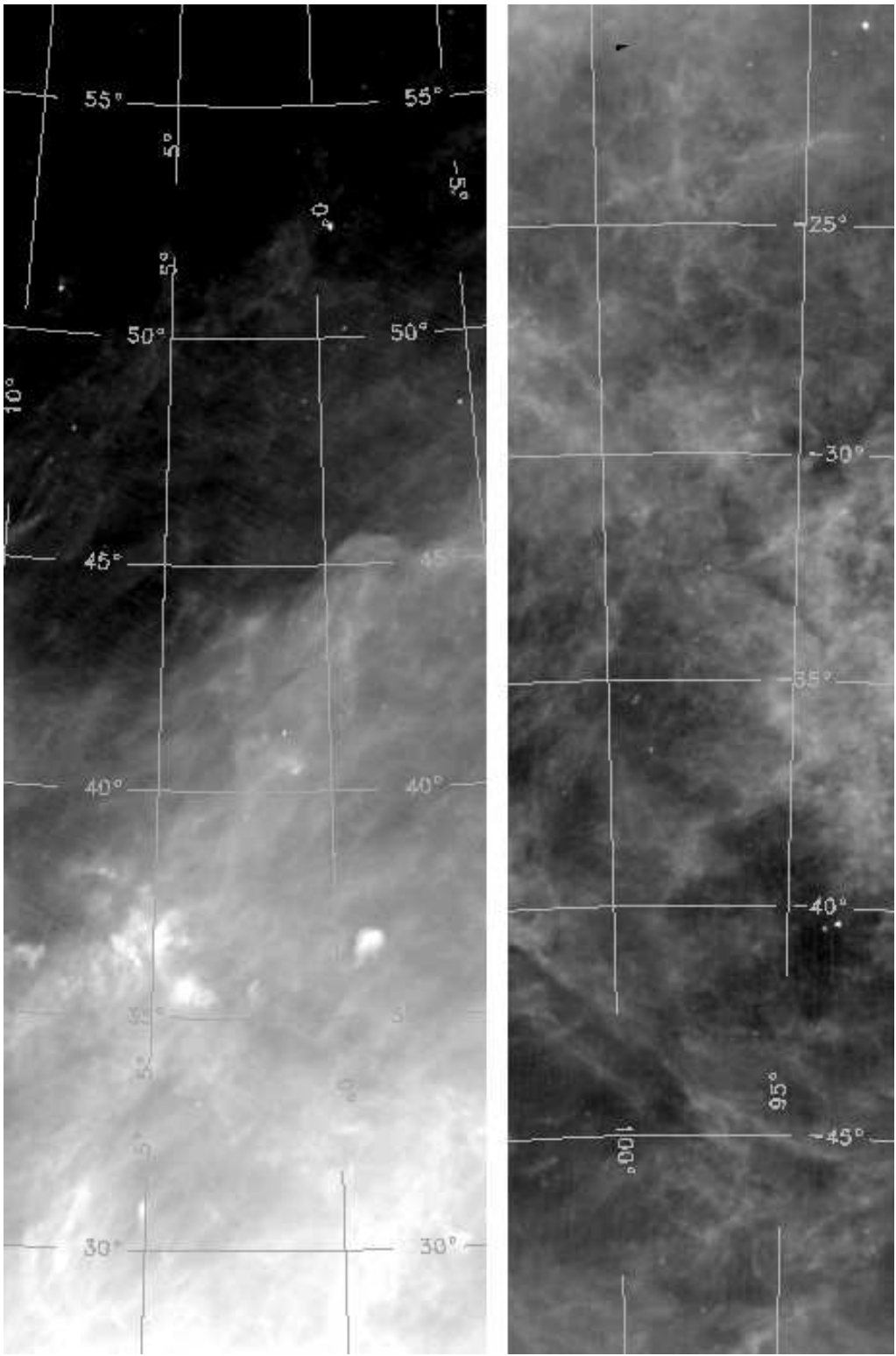}} 
\caption{IRAS $100\,\mu$m images of GWF fields (a), left, and (b), 
right. The brightest regions (saturated) of field (a) have surface 
brightnesses between $25$ and $30\,$MJy/sr. The surface 
brightness decreases to $11\,$MJy/sr at $40^{\circ}$. It is of 
respectively $\sim3\,$MJy/sr and $\sim 1\,$MJy/sr in the brightest and 
darkest parts above $45^{\circ}$. The brightest regions in field (b) 
have $\sim 8\,$MJy/sr surface brightness. The surface brightness falls 
to $5\,$MJy/sr in regions where emission is still discernable and $1.5\,$MJy/sr in the darkest parts. 
Both fields can be seen to be very structured.} 
\label{gordiras}
\end{figure}
\begin{figure}
\resizebox{0.8\columnwidth }{!}{\includegraphics{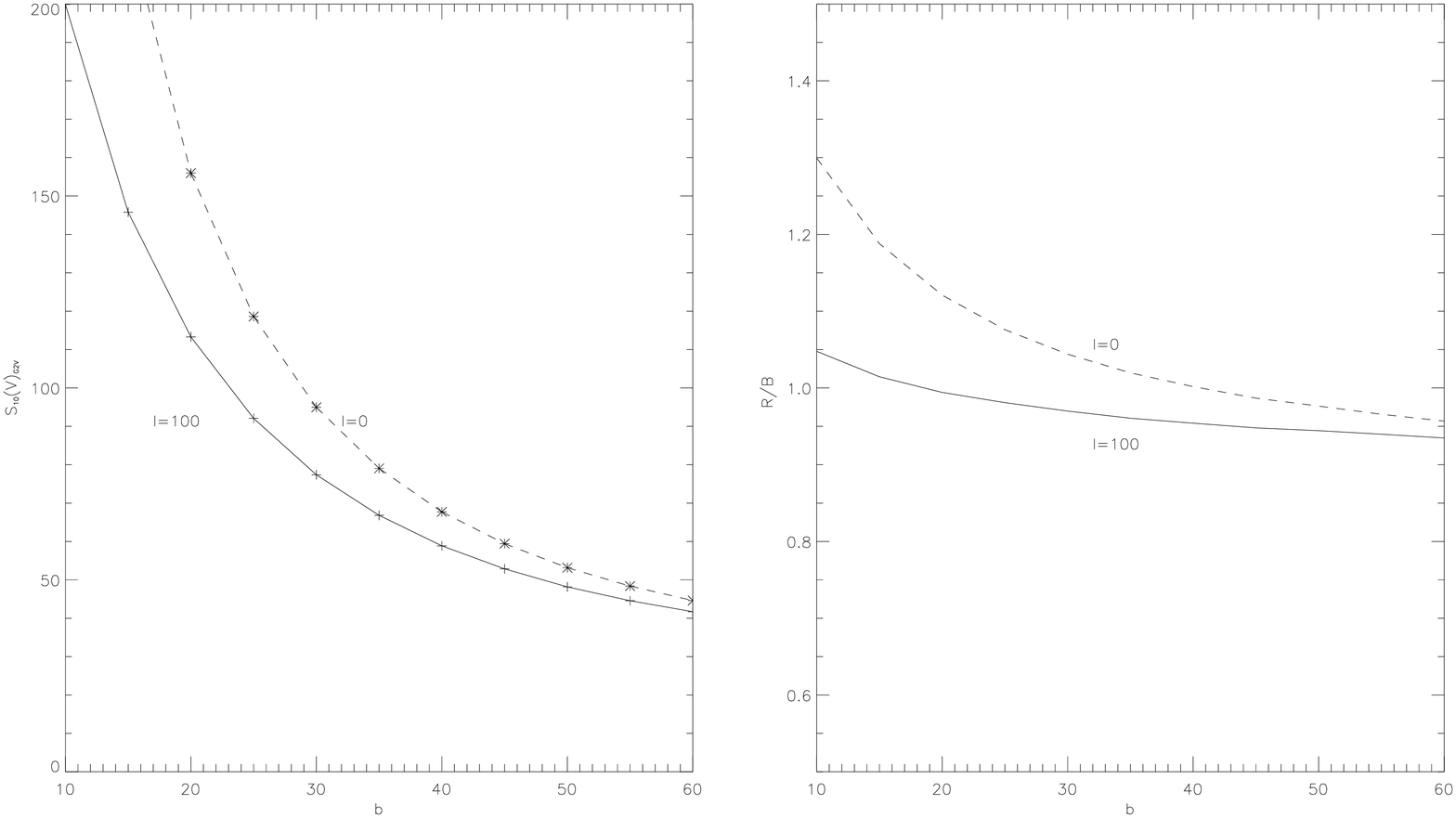}} 
\caption{\emph{left}: Profiles of stars with mag $>6$ surface brightness at $l=0$ and 
$l=100$, for the $B$ band. Computed from Besan\c{c}on Galactic 
Model. \emph{right}: $R/B$ color of the emission.} 
\label{isrffig}
\end{figure}
 \section{Data} \label{data}
To compare the color of DGL and of direct 
starlight, GWF have determined the absolute emissions in the $B$ and 
$R$ bands in two regions of the sky and calculated the $R/B$ ratio of the emissions. 
This method is difficult to apply to groundbased observations since it requires the substraction of all 
foreground (zodiacal light and scattering in the earth's 
atmosphere, see Leinert et al. \cite{leinert98} for a review) sky emissions. 
The difficulty is increased by the low level of emission of the diffuse interstellar medium at 
high galactic latitude. 
Any such attempt should use space observations and substract the direct 
emission of the stars, which can be evaluated from star catalogs. 
Toller (\cite{toller81}, see also Leinart et al.) did this work in 192 regions of the sky using Pioneer measurements.

In boxes $5^{\circ}\times 5^{\circ}$ large, and  
along two cuts at galactic longitude, $l=0$ and $l=100$ (region (a) and 
region (b), figure~\ref{gordiras}), 
GWF have separated the respective $B$ and $R$ emissions 
of bright stars, stars with $\rm mag>6$, and the DGL. 

Bright stars with $m_V<6.5$ were removed during the Pioneer data reduction. 
The remaining emission, comprising stars of $m_V>6.5$, galaxies, and DGL, 
corresponds to the Pioneer curves of the figures~\ref{gord12}, \ref{gord13} 
and  \ref{gord14}. 

The ISGL curves correspond to the expected 
emission of stars and galaxies with magnitude greater than 6.5. 
They are calculated from the GWF `Master Catalog' which is a compilation of different star 
catalogs. 

Not all stars have a $B$ and an $R$ measured magnitude. To 
determine the $R$ and $B$ magnitudes of a star with only a $V$ 
measurement, GWF use an estimate of the $A_V$ 
value in the star direction. 
Determination of $A_V$ assumes a reddening proportional to the 
star distance. 

Figure~\ref{gord12} gives the Pioneer surface brightnesses $P_B$ and 
$P_R$, and the ISGL integrated intensities per 
unit solid angle, $isgl_B$ and $isgl_R$, as measured from earth, for the $B$ 
and $R$ bands, along cuts (a) and (b). 
 \section{The GWF's argumentation} \label{gwfarg}
 \subsection{The GWF analysis of figures~\ref{gord12}, \ref{gord13} and 
 \ref{gord14}} \label{gwf}
GWF's analysis of the curves, figures~\ref{gord12}, \ref{gord13} and 
\ref{gord14}, is short enough to be 
reproduced here: ``From Figure 14, it is obvious that the diffuse ISM 
is redder (larger $PR/PB$ ratio) than either the Pioneer measurements 
or the ISGL. As the scattered component of the diffuse ISM (DGL) is 
bluer (see \S~4.1) than Pioneer measurements, this requires that the 
nonscattered component is present in the red diffuse ISM intensity.
(GWF, p.531)''

It is difficult to accept GWF's premise at face value.
Why should the scattered component of the diffuse ISM (DGL) be
bluer than Pioneer measurements or than the ISGL? 
Where in GWF has it been proven that DGL must be bluer than Pioneer measurements?

In GWF's introduction it is ascerted that: ``The DGL will have a 
bluer color than the integrated starlight because scattering is more 
efficient at shorter wavelengths [in the blue than in the red] (GWF 
p.523)''.
The following sentence states:
``So, if the diffuse ISM color (red/blue ratio) is as red as or redder 
than the integrated starlight and other sources of excess red light 
can be positively excluded, ERE is present (GWF p.523)''.

Note that the first of these GWF assumptions is evidently restricted to low 
column density mediums since an increase of column density will see all blue 
starlight absorbed before red light and the scattered light color turn to 
the red.
If GWF had related the two sentences to suggest that the 
source of the scattered light in one direction is the background 
starlight in that direction, I would fully agree with this assumption since 
interstellar grains are known to scatter light preferentially in the 
forward direction.
The elimination of the 
contribution of the bright stars ($mag<6.5$) may not modify the ISGL 
color enough to invalidate the calculations. This is justified, for 
instance, if the brightest stars are close to the sun or by the small 
number of bright stars at high latitude.

GWF don't seem to make this connection between DGL and background 
starlight.
Their approach of the problem is to calculate what should be the 
scattered starlight in each direction, taking into account the transfer of starlight 
from all directions and throughout the whole galaxy.
 \subsection{The `WP' model} \label{wp}
 \subsubsection{Description of the WP model} \label{wpdes}
\S~4.1 of GWF -where the reader is supposed to find the demonstration 
that the scattered-DGL should have a bluer color than Pioneer's and ISGL's-
is dedicated to the comparison of the Witt-Petersson (`WP') model with the data. 
According to GWF, the WP model represents the galaxy as a `gigantic reflection 
nebula'. 
Calculation of the radiative transfer through the nebula will give a 
model of the scattered light. 
This scattered-DGL model will be compared to the observations.

In order to compute the radiative transfer of starlight through the `gigantic 
nebula' the structure of the medium in 
which light is scattered and the 
radiation field at each position of the galaxy need to be defined.
The result given by the simulation of the radiative transfer through 
the WP model gives an expected radiation 
field for all directions of the galaxy. 

The WP model of the galaxy is constructed from HI surveys.
For each direction of space, the HI column density is converted into 
dust optical depth for the $B$ and $R$ bands by 
multiplication with an appropriate factor. 
The total column density is also divided into interstellar clouds with a spectrum 
of sizes and optical depths given by Witt et al. (\cite{witt97}).

The radiation field GWF use is deduced from 
Pioneer data, with the bright stars re-integrated.
It comprises the DGL. 
 \subsubsection{Results} \label{wpres}
Concerning field (a), the model can hardly fit the data in the $B$ 
band, where scattering is supposed to be the only process involved.
For field (b), whatever grain parameters are chosen, the model is a factor 
of 2 to 4 above the data, which GWF interpret as changes in grain 
properties with longitude.

Despite the fact that the WP model does not fit the data when it 
should, the conclusion of GWF is that the $PR/PB$ 
ratio is a factor of 2 over the WP model expectation. 
Since $PB$ contains starlight and scattered starlight, it is concluded that 
additional red emission must arise in the diffuse interstellar medium.

 \section{Some problems with the GWF argument} \label{crit}
 \subsection{Remarks on the `WP' model} \label{wpcrit}
Not surprisingly, GWF's model does not concur with the 
observations and as such the following questions arise:

What if the `WP model' is an inappropriate one on which to model the galaxy? 

Can the WP model give  a representation 
of the galaxy with enough accuracy to justify the radiative 
transfer calculations?
When the GWF model doesn't fit the data, why is the discrepancy attributed to 
the variations of grain properties and not to the inadequacy of the model?

Why is it necessary to have a representation of the whole galaxy to 
model the DGL in two limited regions? 
Why isn't the model restricted to fields (a) and (b)? Grains are 
known to strongly forward scatter starlight, which implies that 
most of the DGL we receive from one direction originates from 
the scattering of the light of the stars close to that direction.

The radiation field which GWF use in their Monte Carlo simulation of 
the radiative transfer through the Witt representation of the galaxy 
includes DGL which should be the result of the simulation. 
Why isn't direct starlight, which GWF have estimated in their `Master 
Catalog' the input radiation field of the simulation?

GWF never proved that the observed DGL should 
be bluer than the ISGL. Only Monte Carlo simulations with the WP model 
as a representation of the galaxy and an arbitrary radiation field give 
a theoretical color of the scattered starlight 
bluer than that of the ISGL.

Is it necessary to introduce any model at all?
The search for ERE can be restricted to a comparison of the observed color of the 
ISGL (corrected for reddening) and of the DGL in each of the 
$5^{\circ}\times 5^{\circ}$ areas.

In regard to the method and the uncertainties which accompany this model 
one can at best find GWF conclusions doubtful. 
 \subsection{Is the DGL redder than the ISGL?} \label{colrel}
The high values of $S_R/S_B$ in figure~\ref{gord14}, figure~14 in GWF, are due to 
the considerable amplification of the error by the successive
substractions and divisions which led to the calculation of the DGL 
color.

The problem introduced by the amplification of the error through 
these operations can be overcome if it is remarked that 
$S_R/S_B>isgl_R/isgl_B$ is equivalent to 
$P_R/P_B>isgl_R/isgl_B$ since $P_R=S_R+isgl_R$ and $P_B=S_B+isgl_B$.
Hence, comparison of the DGL color and 
the ISGL color can be reduced to a comparison of the Pioneer and the 
ISGL colors, which can be done with much better accuracy.

In both fields the $2$ curves $P_R/P_B$ and $isgl_R/isgl_B$ are very 
close to each other.

Concerning field (b), we have within the error margin:
$P_R/P_B=isgl_R/isgl_B$, which would 
yield: 
\begin{equation}
P_R/P_B=isgl_R/isgl_B=S_R/S_B    
    \label{eq:col}
\end{equation}
Except for the two lower latitude points, the same remark applies to 
field (a). The redder color of the Pioneer data at the two low latitude 
points will be interpreted in section~\ref{exp}.

\section{Relations between the colors of DGL, 
 ISGL and Pioneer} \label{rad}
\subsection{IRAS data} \label{iras}
\subsubsection{The diffuse interstellar medium and IRAS images} \label{iras1}
Because grains scatter starlight in the forward direction, it will be 
considered that DGL in one direction is the light of background stars 
in the same direction scattered by foreground dust.
This is also justified by the large areas which GWF have considered.

It is a wide-spread idea that the diffuse galactic scattered light is 
due to a diffuse interstellar medium, which the WP model attempts to 
represent. 
But the large grains which scatter starlight in the visible have a 
thermal emission, and, considering the importance of the scattering, 
must be detected on the IRAS images.
Therefore, the diffuse medium which, according to GWF, is responsible for 
the DGL in fields (a) and (b)
can be identified to the infrared cirrus shown in figure~\ref{gordiras}.
The GWF areas are in approximate correspondence with the $5^{\circ}\times 5^{\circ}$ 
rectangles in each of the images.

These areas sample a medium with evident structure. 
It is this medium that the WP model pretends to reconstruct from 
its HI emission.
\subsubsection{IRAS images of fields (a) and (b)} \label{iras2}
In all probability the mediums which compose the cirrus of field (b) 
(figure~\ref{gordiras}, right) 
has similar properties in each of the GWF areas.
The decrease of the $100\,\mu$m surface brightness with absolute 
latitude in field (b) is associated with the decrease of the radiation field.
At high absolute latitude, the IRAS surface brightness varies from 
$5$~MJy/sr to $1.5$~MJy/sr. 
With an $100\,\mu$m to visible extinction ratio $I_{100}/A_V$ of 
$18$~MJy/sr/mag (Boulanger~\& 
P\'erault \cite{bou88}), $A_V$ is less than $0.2$ on the average.

The areas of field (a) with $b>45^{\circ}$ have similar 
properties as field (b). They have a low $100\,\mu$m surface 
brightness, a small $A_V$ (on average), and a clear small scale structure.
These areas are the outermost parts of the HI loop at the edge of the 
Scorpio Centaurus region. 

The (a) areas at lower latitude are in the densest parts of the HI loop and have high surface 
brightnesses. Zeta Oph ($m_V=2.6$), a few degrees apart, may 
contribute to the 
heating of the region and to the enhancement, up to $30$~MJy/sr, of 
its infrared emission. In the low latitude areas there is an increase 
of the average HI column density (de Geus \cite{degeus88}), hence of 
the visible extinction.
CO is detected at the brightest IRAS positions (Laureijs et al. 
\cite{laureijs95}).

GWF data can be separated in two. The field (b) and the high latitude 
regions of field (a) have a low column density and a low visible extinction 
on average. The low latitude points of field (a) are clearly a 
different kind of medium with much higher column density and visible extinction.
\subsection{The color of the ISGL} \label{isgl}
In regions where there is interstellar matter, the direct 
starlight is reddened. 
In these regions the color of the direct starlight, $isgl_R/isgl_B$, 
is redder than $\sigma_R/\sigma_B$, 
where $\sigma_R$ and $\sigma_B$ are the surface brightnesses of direct 
starlight corrected for interstellar reddening.

Let $f$ be the filling factor of interstellar matter in one of the 
areas GWF have considered. Assume that the `diffuse interstellar matter' 
has an average low $A_V\sim \tau_V$ value.
This is true in region (b) and in the high latitude regions of (a).

We have for the regions with low $100\,\mu$m emission:
\begin{eqnarray}
    isgl_R &\,=\, & (1-f)\sigma_R+(1-\tau_R)f\sigma_R
    \nonumber \\
     & \,=\,  & \sigma_R(1-f\tau_R)
    \label{eq:isglr}  \\
    isgl_B &\,=\, & \sigma_B(1-f\tau_B),
    \label{eq:isglb}
\end{eqnarray}¥
and
\begin{eqnarray}
    \frac{isgl_R}{isgl_B} &\,=\, &
    \frac{\sigma_R}{\sigma_B}
    \frac{1-f\tau_R}{1-f\tau_B}
    \nonumber \\
    & \,\sim \,  &\frac{\sigma_R}{\sigma_B} (1+f(\tau_B-\tau_R))
        \label{eq:isglcol}
\end{eqnarray}¥
Expressed as a function 
of $A_V$ (Cardelli et al \cite{cardelli89}), equation~\ref{eq:isglcol} 
takes the simple form:
\begin{equation}
        \frac{isgl_R}{isgl_B} \,=\,  \frac{\sigma_R}{\sigma_B} 
        (1+0.6fA_V)
        \label{eq:isglcolr}
\end{equation}¥
The measured color of the stars is of course redder than 
$\sigma_R/\sigma_B$, but the change of color will not be important.
The reddening of direct starlight, $1+0.6fA_V$, for a medium of 
$I_{100}\sim 2$~MJy/sr and $A_V\sim 0.1$ will be of order $1+0.05f$.
Within the margin of error estimated by GWF for the ISGL, $\sim0.1$, 
we can adopt $ isgl_R/isgl_B=\sigma_R/\sigma_B$.

Local increases of column density, due to the small scale structure of 
the medium, will not modify this approximation.
The high resolution images of MCLD123.5+24.9 presented in Zagury et 
al. (\cite{zagury99}) shows the 
presence of high density clumps with a small surface coverage ($\sim 
5'\times5'$). In field (b) and for the (a)-field regions where 
$b>40^{\circ}$, high density clumps must occupy only a small fraction 
of the surface.  
In such clumps the number of stars diminishes with $A_V$ and 
the color of ISGL in the lower column density medium, which probably occupies most of 
the volume in regions (a) and (b), will determine the ISGL color of 
the $5^{\circ}\times 5^{\circ}$ areas.

The result does not hold for the low latitude points of field (a). 
There is a net increase of the average column density which can be 
evaluated by: 
\begin{eqnarray}
 \frac{isgl_R}{isgl_B}&\,=\,&\frac{\sigma_R}{\sigma_B} e^{\tau_B-\tau_R} 
 \nonumber \\
 &\,=\,&\frac{\sigma_R}{\sigma_B} e^{0.5A_V}
 \label{eq:isglgred}
\end{eqnarray}
While the reddening is, as seen before, small for small $A_V$ values, 
for $A_V=0.5$ the ISGL is $1.3$ times redder than the stars' color.

The latter remarks also show that GWF's 
estimate of the $B$ and $R$ magnitudes of the stars may have been biased when
the $A_V$ value of the stars were used to construct the `Master Catalog'.
To estimate $A_V$, GWF assume an average extinction of $0.6$~mag/kpc.
This approximation applies only if the interstellar 
matter in the star direction has a low column density.
In the case of larger column densities, the GWF estimate of the 
color of starlight will be bluer than it is in reality.

The distance to the cirrus of field (a) should be 
$200$~pc at most, since it belongs to the Scorpio Centaurus 
region. 
For an $A_V$-value of 0.5 for instance, the extinction of all stars at closer 
distance than $300$~pc will be underestimated.
The effect increases with column density and will be 
more pronounced in the low latitude regions 
of field (a).
It will affect the $B$ band more than the $R$ band since the 
$E(B-V)/A_V$ coefficient used by GWF is twice $E(V-R)/A_V$.
The color of the ISGL estimated by GWF from the Master catalog will be bluer 
than it should.
It may explain the drop of the ISGL color of the lowest latitude 
point of field (a), figure~\ref{gord14}.
\subsection{The Pioneer color} \label{pioneer}
The Pioneer surface brightness comprises the ISGL ($isgl \sim \sigma$) and the 
DGL.

If the scattering volume is a medium of low optical 
detph $\tau_B$ in the $B$ band, the DGL surface brightness will be at most: 
$\omega \tau_B$, where 
$\omega$ is the albedo, assumed to be constant ($\sim0.6$) at optical 
wavelengths. 
Use of equations~\ref{eq:isglr} and \ref{eq:isglb} gives:
\begin{eqnarray}
    P_R \,=\,\sigma_R(1- f(1-\omega) \tau_R)
    \label{pb}    \\
 P_B \,=\,\sigma_B(1- f(1-\omega) \tau_B)
    \label{pb1}   
\end{eqnarray}¥
Then:
\begin{eqnarray}
   \frac{ P_R}{P_B} & \,=\,& \frac{\sigma_R}{\sigma_B}
   \frac{(1- f(1-\omega) \tau_R)}{(1- f(1-\omega) \tau_B)}
    \nonumber    \\
    & \,\sim\,& 
   \frac{\sigma_R}{\sigma_B}(1+f(1-\omega)(\tau_B-\tau_R))
   \label{eq:pred}\\
   \frac{ P_R}{P_B} & \,=\,& 
   \frac{isgl_R}{isgl_B}(1-\omega(\tau_B-\tau_R))
   \label{eq:pred1}
\end{eqnarray}¥
The Pioneer color is in between the ISGL and the DGL colors. 
Equation~\ref{eq:pred} can be simplified with $\omega\sim0.6$ and Cardelli, Clayton and Mathis 
(\cite{cardelli89}) relations:
\begin{eqnarray}
     \frac{ P_R}{P_B} & \,=\,& \frac{\sigma_R}{\sigma_B}(1+0.3fA_V) 
    \label{eq:preds} \\
     \frac{ P_R}{P_B} & \,=\,& \frac{isgl_R}{isgl_B}(1-0.15fA_V)  
    \label{eq:preds1}
\end{eqnarray}¥
Equations~\ref{eq:preds} and  \ref{eq:preds1} show that within GWF 
error margin equality between $P_R/P_B$, $isgl_R/isgl_B$, 
$\sigma_R/\sigma_B$ and $dgl_R/dgl_B$, will be satisfied for mediums of low column density.

\subsection{Effect of the cirrus small scale structure on the observed 
colors} \label{sss}
Small scale structure affects the surface brightness of starlight.
Stars with no interstellar matter on their line of sight will have little or 
no reddening, while stars behind a cirrus are reddened.
The clumpiness of the regions GWF have chosen, revealed by the IRAS 
images, will certainly affect the evaluation of the starlight 
surface brightness since GWF have considered an average reddening by 
unit distance (section~\ref{data}) for all stars, regardless of the 
increase of interstellar matter in some directions.
As pointed out in section~\ref{isgl}, the color of direct starlight may be redder 
than estimated by GWF.

Small scale structure also modifies the color of the DGL since small clumps of different 
$A_V$ can be mixed in the beam.
Even if the optical depth averaged over the beam of the observations 
is small, the existence of clumps of higher $A_V$ than average cannot be discarded.
These clumps will redden the color of the DGL from the color of a low 
column density medium. 
Regions of higher than average $A_V$, such as the low latitude areas of 
field (a), are more likely to show this effect.

These effects of the small scale structure of the interstellar medium on GWF 
data cannot be totally leaved over.
Each of the GWF areas is large compared to the size at which 
the interstellar medium is stuctured.
IRAS images show that this size is less than a few arcminutes.
Higher resolution observations (Falgarone et al. \cite{falgarone}, Zagury et al. 
\cite{zagury99}) give sizes of a few arcseconds at most for the 
entities which compose the interstellar medium. 
\subsection{The comparison of Pioneer and ISGL colors} \label{colcomp}
The preceding sections show that a precise comparison of Pioneer and 
direct starlight data will sharply 
depend on the reliability of the GWF master catalog and on the cirrus 
structure.

For the GWF regions of low average optical depth, plot~\ref{gord14} 
is in accordance, within the error margin given by GWF, with equality~\ref{eq:col}.

Concerning the two lower latitude point of the field (a) 
different reasons may explain the drop in ISGL color.
These points are different from the others since the region  has a much 
higher $A_V$ and cannot be considered as a `diffuse medium'.
This was not taken into account by GWF for the 
estimate of the direct starlight $B$ and $R$ magnitudes.
The medium also contains regions with more extinction which may modify 
the DGL color and redden the Pioneer color.
 \section{A qualitative comparison of IRAS image and the GWF data} 
 \label{exp}
A remarkable difference between fields (a) and (b) can be seen in
figure~\ref{gord12}. In 
field (b), $isgl_R$ and $isgl_B$ decrease with 
increasing (absolute) latitude, as is
expected. The $B$ surface brightness follows a $\sim 
1/\sin |b|$ law from 
$b=-28$ 
to $b=-55$, with a higher value than expected at $b=-22$. It is 
slightly under but close to the prediction of the Besan\c{c}on Galactic 
Model (figure~\ref{isrffig}). 
The $R/B$ color  closely follows the model.

In field (a) the ISGL is nearly constant and $isgl_R$ and $isgl_B$ have parallel variations. 
Low latitude points in (a) have lower ISGL than in (b) and than what 
is predicted ($\rm \sim 75\,S_{10}(V)_{G2V}$) by the Besan\c{c}on Galactic Model, 
figure~\ref{isrffig}, while 
higher latitude points ($|b| >45^{\circ}$) have the same values at both 
longitudes. 
The ratio of the DGL 
to the total Pioneer emission in the $R$ band, $S_R/P_R$ (calculated 
from  figure~\ref{gord12} and figure~\ref{gord13}) is between $0.45$ and 
$0.5$ for the $3$ lowest latitude points in (a), while for the higher 
latitudes points and for all points in (b), it is between $0.2$ and $0.3$. 

The lower than expected starlight 
emission  for field (a) low latitude points, 
along with the relative increase of DGL $R$ emission, 
are easily interpreted as extinction effects due to 
the average increase of interstellar matter along the 
line of sight (section~\ref{iras}). 
The increase of the column density increases the extinction of starlight.

Corresponding to this extinction of starlight, there is a sharp rise in the DGL
emission between the 3 (a) points at latitude $42.5$, $37.5$ 
and $32.5$, figure~\ref{gord13}. The $R$ emission between latitudes $37.5$ and $32.5$ reaches 
a ceiling which corresponds to a decrease of $B$ emission, $S_B$. This 
can also be understood by the average increase 
of column density: absorption starts to dominate scattering in the $B$ band and 
the $R$ surface brightness is increased.

For the two low latitude points of field (a) in figure~\ref{gord14} 
departure from relation~\ref{eq:col} seems certain.
In this region we are clearly outside the low column density 
approximation, supposed by GWF, hence outside the framework of their study.
The understanding of the ISGL color in this region deserves further 
investigations but two reasons may contribute to the blue color of the ISGL.
One reason is the GWF method to estimate the ISGL magnitude which 
applies to low column density mediums only (section~\ref{isgl}).
The second reason is if  Zeta Oph, a few degrees apart, participates in the illumination of 
the region. 
Note that illumination by Zeta Oph ($m_V=2.6$) on such a large 
scale may be difficult to justify. 

The correlation between the IRAS images of both fields 
(figure~\ref{gordiras}) 
and the DGL $R$ emission (figure~\ref{gord13})  makes it likely  that 
most of the emission measured by Pioneer comes from the infrared 
cirrus. 
In the visible, the cirrus scatter the light of background stars.
The relation between the color of the radiation field due to the stars 
and the color of the DGL is given by equation~\ref{eq:col}. 

The DGL optical emission at low latitude in field (a) may in 
part be due to illumination by the star Zeta Oph, a few degrees apart.

 \section{Conclusion}
GWF have compared the color of the diffuse galactic light at high 
galactic latitude, deduced from observations of the Pioneer 
satellite, to the color of the light scattered by a diffuse 
medium with a certain cloud size distribution. 
GWF conclude that ERE is required to explain the 
amount of diffuse galactic emission at high latitude and can represent up to $50 \% $ 
of this emission.

In GWF it is said that, ``An accurate calculation of the DGL should 
include the effects of multiple scattering, the cloudiness of the 
interstellar medium, and the observed anisotropy of the 
illuminating radiation field.''
It should also include a reliable description of the interstellar 
medium.
It would be very surprising if the WP model or any other, as sophisticated 
as it might be, can deduce from 
the measure of the HI emission in one direction the organisation of 
the interstellar matter in that direction. 
It is also surprising that when the WP model does not match the data it 
is supposed to match, changes in the properties of the 
interstellar grains are invoked rather than the validity of the model. 
It seems more plausible that observations do not agree with the WP 
model because this model does not properly describe the reality of the 
interstellar medium. 

The exact structure of the interstellar medium the WP model should 
reproduce is shown at a scale of $2'$ by the IRAS $100\,\mu$m images.
These images reveal the thermal emission of the same large grains 
responsible for the scattering of background starlight in the visible.
They would have been a better basis to model the interstellar medium 
than the HI emission used in GWF.
This medium may be of low column density on the 
average but is extremely structured and may have local column density 
enhancements.
Some regions, the lower latitude points of field (a), have a clear 
increase of their average column density.
It is very unlikely that the WP model reproduces this medium with the 
accuracy necessary to justify their conclusions.

The uncertainty of the DGL color as calculated in GWF is extremely large compared to 
the errors on Pioneer integrated light color or on the 
direct starlight color. 
Comparison of the  DGL and the starlight colors should be replaced by a 
comparison of  the starlight to the Pioneer colors, both of which are 
determined with less relative error.

Within the error margin given in GWF, there is equality between the 
color of the integrated starlight and the color given by the Pioneer 
measurements.
It implies equality with the color of the DGL.
There may be a tendency for the starlight surface brightness 
calculated by GWF to be slightly bluer than Pioneer integrated light.
This tendency correlates with mediums of higher $A_V$ and can be 
attributed to two effects.
Starlight is redder than estimated by GWF who assume a low and 
equal reddening in all directions.
It modifies the GWF estimation of the $R$ and/or $B$ magnitude of the 
stars.
Small scale structure will redden the color of DGL and of 
Pioneer from the color of a low column density medium.

A better understanding of the optical emission in the high 
latitude directions comes from the consideration that the DGL in these 
directions is the light of background stars scattered by the nearby 
infrared cirrus.
For the GWF fields (a) and (b), there are relations
between the cirrus  observed on the $100\,\mu$m IRAS image, 
the visible emission of background 
stars, and the DGL.
The IRAS $100\,\mu$m emission in the GWF field (b) and the high latitude 
regions of field (a) attests to low column 
densities in the average.
The corresponding DGL surface brightness is low.
Increase of the column density of dust in the low latitude areas of 
field (a) attenuates starlight and 
increases the $R$ surface brightness of the DGL, while the increase 
in the $B$ surface brightness is limited because of absorption. 

ERE is not needed to 
explain the DGL at high galactic latitude (this paper) or in bright 
nebulae (Zagury, `Is there ERE in bright nebulae?', submitted). 
Nor is it needed to explain the emission in the Red Rectangle 
(Zagury, in preparation), the milestone of ERE. 
There might 
be no ERE at all in interstellar space, which can be inferred from 
present day observations.


\begin{thebibliography}{}
       
   \bibitem[1978]{bol} Bolhin R.C., Savage B.D, Drake J.F, 1978, ApJ, 
224, 132

   \bibitem[1988]{bou88} Boulanger F., P\'erault M., 1988, ApJ, 330, 964

   \bibitem[1989]{cardelli89} Cardelli J.A., Clayton, Mathis J.S., 
1989, ApJ, 345, 245

   \bibitem[1988]{degeus88} de Geus E., 1988, Ph.D. thesis, Leiden 
   University

   \bibitem[1998]{falgarone}Falgarone E., et al., 1998, A\&A, 331, 669

   \bibitem[1998]{gordon} Gordon K.D., Witt A.N., Friedmann B.C., 
1998, ApJ, 498, 522

   \bibitem[1994]{gu2} Guhathakurta P., Cutri R.M., 1994, ASP Conf. Series, 58, 34

   \bibitem[1941]{henyey} Henyey L.C., Greenstein J.L.,  1941, ApJ, 93, 70
   
   \bibitem[1995]{laureijs95} Laureijs R.J., et al., 1995, ApJS, 101, 
   87

   \bibitem[1998]{leinert98} Leinert Ch., et al., 1998, A\&AS, 127, 1

   \bibitem[1962]{lynds62} Lynds B.T., 1962, ApJS, 7, 1 
   
    \bibitem[1986]{witt86} Witt A.N., Schild R.E, 1986, ApJS, 62, 839

   \bibitem[1976]{sandage} Sandage A., 1976, AJ, 81, 954 

   \bibitem[1936]{struve36} Struve O., Elvey C.T., 1936, ApJ, 83, 162 

   \bibitem[1981]{toller81} Toller G., 1981, Ph.D. thesis, State University 
   of New York at Stony Brook

   \bibitem[1985]{devries85} de~Vries C.P., Le Poole R.S., 1985, A\&A, 
   145, L7

   \bibitem[1999]{zagury99} Zagury F., Boulanger F., Banchet V., 1999, 
     A\&A, 352, 645 (paper~I)

   \bibitem[1997]{witt97} Witt A.N., Friedmann B.C., Sasseen T.P, 1997, 
   ApJ, 481, 809

\end{thebibliography}
\end{document}